\documentclass[prl,amssymb,twocolumn,tightenlines,showpacs,superscriptaddress]{revtex4-1}

\usepackage{editing}

\usepackage{units}
\usepackage{graphicx}
\usepackage{dcolumn}
\usepackage{bm}
\usepackage{color}
\usepackage{hyperref}
\hypersetup{colorlinks, citecolor=blue, linkcolor=blue, urlcolor=blue}

\usepackage{amsmath}
\usepackage{amssymb}
\usepackage{gensymb}

\usepackage{tikz}
\usepackage{standalone}

\usepackage{booktabs} 
\usepackage{colortbl} 
\usepackage{xcolor} 
\usepackage{xfrac}
\usepackage{textcomp}
\usepackage{xspace}

\usetikzlibrary{decorations.pathmorphing}
\usetikzlibrary{decorations.markings}
\usetikzlibrary{arrows,decorations}
\usetikzlibrary{shapes.misc, positioning}
\usetikzlibrary{shapes.arrows, fadings, calc}

\newcommand{\ket}[1] {| #1 \rangle}

\newcommand{\Ca}{\ensuremath{^{40}{\rm Ca}^+}\xspace}

\newchange{}{DarkGreen}{OrangeRed}%
\changeOnOff

\begin{document}


\title{Improved Test of Local Lorentz Invariance from a Deterministic Preparation of Entangled States}

\author{Eli Megidish}
\author{Joseph Broz} \author{Nicole Greene} \author{Hartmut H\"affner}
\affiliation{
Department of Physics, University of California, Berkeley, California 94720, USA
}

\begin{abstract}
The high degree of control available over individual atoms enables precision tests of fundamental physical concepts. 
In this Letter, we experimentally study how precision measurements can be improved by preparing entangled states immune to the dominant source of decoherence. Using \Ca ions, we explicitly demonstrate the advantage from entanglement on a precision test of local Lorentz invariance for the electron.
Reaching the quantum projection noise limit set by quantum mechanics, we observe for bipartite entangled states the expected gain of a factor of two in the precision. Under specific conditions, multipartite entangled states may yield substantial further improvements. 
Our measurements improve the previous best limit for local Lorentz invariance of the electron using \Ca ions by factor of two to four to about $5\times10^{-19}$. 
\end{abstract}

\maketitle

Quantum entanglement can be harnessed to enhance the measurement precision beyond the standard quantum limit  \cite{Caves1981,Walls1983,Slusher1985,Yurke1986a,Kitagawa1993,Braunstein1994,Wineland1994}. 
In particular, demonstrations with photons  \cite{Higgins2007,Nagata2007}  and atomic systems \cite{Meyer2001,Leibfried2004}  have
reached the fundamental Heisenberg limit. However, translating these techniques into actual improvements of precision measurements or fundamental tests appears to be difficult.
One reason is that the correlation between the particles, which enhances the precision, is also prone to decoherence; often negating the advantage of entanglement \cite{Huelga1997,Kacprowicz2010,Thomas-Peter2011,Kessler2014}. On the other hand quantum correlations can not only be used to improve the signal, but also to engineer a quantum state insensitive to certain noise sources and still sensitive to the desired quantity thereby improving  high precision measurements \cite{Roos2006,Chwalla2007,Pruttivarasin2015}. 
Probabilistic preparation of entangled states \cite{Chwalla2007} and the resulting metrological improvement have been presented previously \cite{Pruttivarasin2015}.
Here we study how in this case moving from a separable quantum state created by a probabilistic source \cite{Chwalla2007,Pruttivarasin2015} to an entangled state, as used in Ref. \cite{Roos2006}, improves the signal-to-noise ratio of a test of the local Lorentz invariance (LLI) of the electron. 

LLI-violation effects can be classified in the framework of the standard model extension  \cite{Kostelecky2011}. 
In particular, the Lagrangian describing the electron is modified to allow for local Lorentz violations while maintaining all other symmetries.
In the nonrelativistic limit, this can be described by the effective Hamiltonian  \cite{Kostelecky2011,Hohensee2013c,Pruttivarasin2015}:
\begin{equation} \label{energy_shift}
    \delta \mathcal{H}  = -C_0^{(2)} \frac{p^2-3p_z^2}{6m_e}
\end{equation}
where $m_e$, $p$, and $p_z$ are the electron mass, the electron total momentum, and the momentum projection along the quantization axis, respectively. The $C_0^{(2)}$ parameter contains elements of the symmetric, traceless, and frame dependent $C_{MN}$ tensor quantifying the LLI violation~\cite{Hohensee2013c}. 
We use the Sun-centered celestial reference frame (SCCEF) indicated with coordinate indices ($T,X,Y,Z$ ) to uniquely specify the $C_{MN}$ values. 

To date, the most sensitive LLI tests for electrons have been published with Dy atoms \cite{Hohensee2013c} and Ca$^+$ ions \cite{Pruttivarasin2015}.
Theoretical calculations show that similar measurements using Yb$^+$ can improve  the existing bound considerably  \cite{Dzuba2016a} and measurements underway are expected to be published soon by the PTB group at Braunschweig.
Recently, it was also suggested to use dynamical decoupling techniques to suppress magnetic field noise making single ions, neutral atoms and highly charged ions attractive for LLI-violation searches \cite{Shaniv2018}. 


The hypothetical LLI energy shift for the $^2D_{5/2}$ manifold in $^{40}$Ca$^+$ is characterized by \cite{Pruttivarasin2015}
\begin{equation} \label{LLI_shift}
     \nicefrac{E_{LLI}}{h} =C_0^{(2)} [2.16\times10^{15}-7.42\times10^{14} m_J^2] \:{\rm Hz}, 
\end{equation}
where $m_J$ is the projection of the total angular momentum on the magnetic field. For a maximal LLI sensitivity, we  prepare an ion in a superposition of  $\ket{D_{\pm 1/2}}=\ket{^2D_{5/2}, m_J = \pm 1/2}$  and
$\ket{D_{\pm 5/2}}=\ket{^2D_{5/2}, m_J = \pm 5/2}$ parallel and orthogonal to the magnetic field, respectively. 
The orientation of the magnetic field is fixed with respect to the Earth's frame, and hence the rotation of the Earth will rotate the orientation of the electronic wave function with respect to the SCCEF frame. Thus, spacelike hypothetical LLI violations
will modulate the phase between the two amplitudes with 12h and 24h periodicities. 
For single ions, the main sources of electronic decoherence are magnetic field fluctuations. 
To suppress this noise, we use two trapped ions labeled $0$ and $1$ in the state:
\begin{equation} \label{Eq.LLI_state}
    \ket{\psi^{0,1}}= \nicefrac{1}{\sqrt{2}}(\ket{D^0_{5/2},D^1_{-5/2}}+\ket{D^0_{{1/2}},D^1_{-1/2}}).
\end{equation}
This is a decoherence free state with respect to global magnetic field fluctuations because the magnetic moments of both ions point in opposite directions \cite{Roos2006}.

For the experiments, we trap two $^{40}$Ca$^+$ ions in a radio-frequency Paul trap. 
The quantization axis is defined by a permanent magnet generating a magnetic field of $3.72$~G oriented $\sim68^\degree$ east of north.
The most relevant vibration mode of the ion crystal is the axial c.m. mode at $\sim830$~kHz, which is used to entangle the ions.
A $729$~nm narrow linewidth laser light is used to address Zeeman transitions between the $^2S_{1/2}$ and  $^2D_{5/2}$ states.
One $729$~nm beam, aligned along the trap axis, couples to both ions. A perpendicular second beam, termed the local beam, is tightly focused to address only one of the ions with less than $1\%$ laser intensity leakage onto the other ion. The electronic state of the ions is detected by collecting photons scattered from the $^2S_{1/2}\,\rightarrow\,^2P_{1/2}$ transition. 

The state preparation starts with Doppler cooling followed by optical pumping of both ions to the $\ket{S_{-1/2}}=\ket{^2S_{1/2}, m_J=-1/2}$ state using laser light on the
$\ket{S_{1/2},m_J=1/2}\leftrightarrow \ket{D_{5/2},
m_J=-3/2}$ and $\ket{D_{5/2}}\leftrightarrow \ket{P_{3/2}}$ transitions \cite{Haeffner2008review}. 
To achieve high fidelity operations, we further cool the axial c.m. and stretch modes to the ground state using
sideband cooling. 
Finally, using the local beam, we optically pump ion no. 0, using the same scheme as for initialization, into
the opposite spin state preparing the $\ket{S^0_{1/2},S^1_{-1/2}}$ state with a fidelity of $>99\%$ where the superscript indicates the ion and the subscript the magnetic quantum number.  

\begin{figure}
  \includegraphics[angle=0,width=0.45\textwidth]{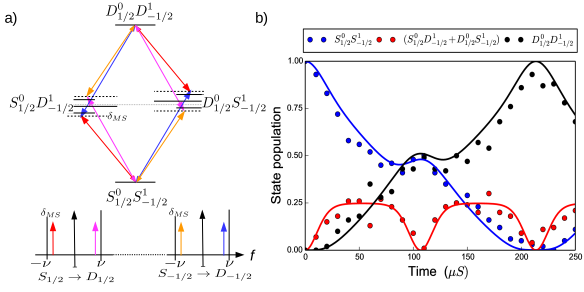}
  \vspace*{-0.25cm}
  \caption{ \label{fig:MS_gate} 
  MS gate scheme. 
  (a) Laser field tones (colored arrows) applied to the ions to generate the MS gate, where black lines represent atomic resonances and motion sidebands transitions .  
  (b) Population dynamics, measured and theory (solid lines), as functions of the pulse duration of the applied laser fields. 
    }
\end{figure}

To entangle the ions prepared in the different spin states, $\ket{S^0_{1/2},S^1_{-1/2}}$, we apply the M\o lmer-S\o rensen (MS) scheme \cite{Molmer1999} in a similar manner as what has been done to entangle different ion species \cite{Tan2015,Ballance2015}. 
In particular, we generate two sets of two bichromatic laser fields, with each centered around the $\ket{S_{1/2}} \rightarrow \ket{D_{1/2}}$ (  \textit{C}3)  and the $\ket{S_{-1/2}} \rightarrow \ket{D_{-1/2}}$ (\textit{C}4) carrier transitions. 
 [see Fig. \ref{fig:MS_gate}(a)]. 
Each set consists of two tones each  ($\delta_{MS} = 10$~kHz) detuned from the c.m. red sideband and blue sideband of the respective carrier transition [Fig.\ref{fig:MS_gate}(a)]. Thus, ion no.0 can only be resonantly excited to the  $\ket{D_{1/2}}$ state if, simultaneously, ion no. 1 is excited to the $\ket{D_{-1/2}}$, and vice versa. 

We measured the population evolution under the applied laser fields [Fig.\ref{fig:MS_gate}(b)].
Both ions are rotated from the initial state to the entangled state:
\begin{equation} \label{Eq_MS_state}
    \ket{S^0_{1/2},S^1_{-1/2}} \rightarrow 
\nicefrac{1}{\sqrt{2}}( 
    \ket{S^0_{1/2},S^1_{-1/2}}+
    \ket{D^0_{1/2},D^1_{-1/2}}
    ).
\end{equation}
Due to the small detuning $\delta_{MS}$ from the sideband transitions, the $\ket{S^0_{1/2}D^1_{-1/2}} and \ket{D^0_{1/2}S^1_{-1/2}}$ states are transiently populated. We carefully adjust the laser powers such that, at the  gate time of $t_g=1/\delta_{MS}=(106\pm5)$~{\textmu}s, the nondesired states are depopulated and an equal superposition is achieved \cite{Roos2008b,Akerman2015}. We further need to take into account that the presence of the four laser tones induces substantial ac-Stark shifts on the respective carrier transitions on the order of 10~kHz \cite{Haeffner2003a}. To counter those, we introduce  frequency offsets for each pair of sideband transitions and adjust them to maximize the gate fidelity. 

To quantify the gate fidelity, we measure the coherence between the two amplitudes in Eq.(\ref{Eq_MS_state}). We apply local $\pi/2$ rotations on the respective carrier transitions and measure the parity \cite{Sackett2000}. 
From the amplitudes of the entangled state before the rotation and the parity fringe amplitude we calculate the gate fidelity to be $94\%$.
The gate fidelity is limited by $\sim5\%$ laser intensity noise, which creates ac-Stark fluctuations leading to phase fluctuations in the entangled state in Eq.(\ref{Eq_MS_state}). We estimate that this reduces the fidelity by $\sim2.5\%$. A large fraction of the remaining infidelity is likely due to laser frequency noise.

\begin{figure}
\includegraphics[angle=0,width=0.45\textwidth]{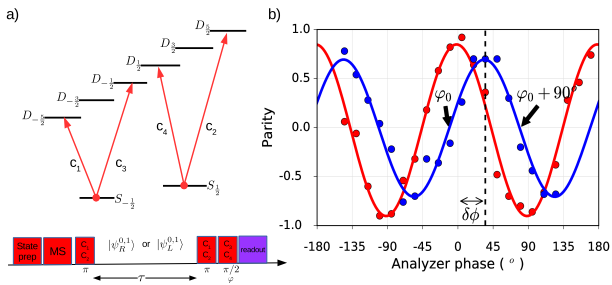}
\caption{
\label{fig:LLI_parity} 
(a) Relevant Ca$^+$ electronic energy levels, transitions, and the experimental sequence for the generation and interrogation of the LLI target state.
(b) Parity, measurements, and fit as a function of the laser field phase $\varphi$ for $5$~ms (red) and $105$~ms (blue) wait times. By measuring the parity at phases $\varphi_0$ and $\varphi_{90}$, we estimate the accumulated phase $\delta \phi$ with respect to the $5$~ms wait time.  
}
\end{figure}

In order to prepare the LLI target state given in Eq.(\ref{Eq_MS_state}), we apply carrier rotations  on the $\ket{S_{\pm1/2}} \rightarrow \ket{D_{\pm5/2}}$ transitions, labeled as \textit{C}1 and \textit{C}2, respectively.
In particular, 
$R_{\textit{C}1}(\pi)$  rotates the $\ket{S_{-1/2}}$ population to $\ket{D_{-5/2}}$ and $R_{\textit{C}2}(\pi)$ rotates the $\ket{S_{1/2}}$ population to $\ket{D_{5/2}}$  [see Fig.~\ref{fig:LLI_parity}(a)]. 

The LLI target state evolves for a duration $\tau$, accumulating a phase $\phi$ between the two amplitudes.
Measuring the phase $\phi$ is accomplished by a parity measurement similar to that used for assessing the quality of the MS gate described before. We first rotate the LLI state back to the $S_{|1/2|},D_{|1/2|}$ subspace using  $R_{\textit{C}1}(\pi)$ and $R_{\textit{C}2}(\pi)$ and then apply $R_{\textit{C}3}(\pi/2,\varphi) and R_{\textit{C}4}(\pi/2,\varphi)$ to interfere the amplitudes; see  Fig.~\ref{fig:LLI_parity}(a). Note that the phases on \textit{C}3 and \textit{C}4 are defined with respect to the phase of the effective carrier transition used in the MS gate. The resulting parity is a
function of the phase $\phi+2\varphi$ between both energy eigenstates and the laser phases according to
\begin{eqnarray} \label{Eq_Parity_LLI}
    P = A{\rm cos}\left(\phi + 2\varphi  \right), \quad 
    \phi ={\Delta E }\tau/{\hbar}= 2\pi f  \tau\: , 
\end{eqnarray}
where $A$ is the interference amplitude.
We calculate the phase $\phi$ from the interference amplitude and by measuring the parity at two consecutive zero crossings of the laser phase labeled  $\varphi = \varphi_0, (\varphi_0+90^\circ)$ [see Fig.\ref{fig:LLI_parity}(b)].
The accumulated phase $\phi$ is proportional to the wait time $\tau$ and the energy difference $\Delta E=hf$ between the $\ket{D^0_{1/2}D^1_{-1/2}}$ and $\ket{D^0_{5/2}D^1_{-5/2}}$ states.

The accuracy of the phase measurement is proportional to the coherence of the state given by the parity fringe contrast. 
For short wait times, we measure a parity fringe contrast of 87\% [Fig. \ref{fig:LLI_parity}(b)]. This contrast is mainly limited by the MS gate fidelity of 94\% as well as the four $\pi$ pulses required to prepare and interfere the LLI state. After cooling the axial modes of the ion string, we achieve about 99\% fidelity for each of the $\pi$ pulses.  After a wait time of  $105$~ms, the contrast of the parity flops decreases  to $70\%$. The reduction, in contrast, is mainly due to spontaneous decay of either of the two ions with a total probability of 17\% (excited state lifetime of $\tau \approx 1.2$~s \cite{Kreuter2005}).
 
Phase drifts in the LLI-state preparation due to changes in the laser intensity or other miscalibrations do not allow for an absolute measurement of the phase. Instead, we remove these phase drifts by interleaved calibrations of the state preparation alternating between wait times of $\tau=5$~ms and $\tau= 105$~ms.  
The effective wait time of 100~ms was chosen to be a multiple of the period of the power grid of 60~Hz to average over a full period, thereby removing systematic effects due to slow variations of the amplitude of the magnetic field variations during a power grid cycle.

The magnetic field of 3.72~G is supplied by a single permanent magnet, which results in a magnetic field gradient of ~0.8mG.
Due to the magnetic field gradient, ion no. 0 would experience an excess Zeeman phase shift.
To remove its effect, we also alternate between states $\ket{\psi^{0,1}_R}$ and $\ket{\psi^{0,1}_L}$: 
\begin{eqnarray} \label{LLI_stateII}
    \ket{\psi^{0,1}_R}= \nicefrac{1}{\sqrt{2}}(\ket{D^0_{5/2},D^1_{-5/2}}+\ket{D^0_{1/2},D^1_{-1/2}}), 
    \nonumber
    \\ 
    \ket{\psi^{0,1}_L}= \nicefrac{1}{\sqrt{2}}(\ket{D^0_{-5/2},D^1_{5/2}}+\ket{D^0_{{-1/2}},D^1_{1/2}}).
\end{eqnarray} 
We measure the frequency $f=\delta \phi/t $  for each state and average the measured frequencies to cancel out the magnetic field gradient accordingly.

\begin{figure} 
\includegraphics[angle=0,width=80mm]{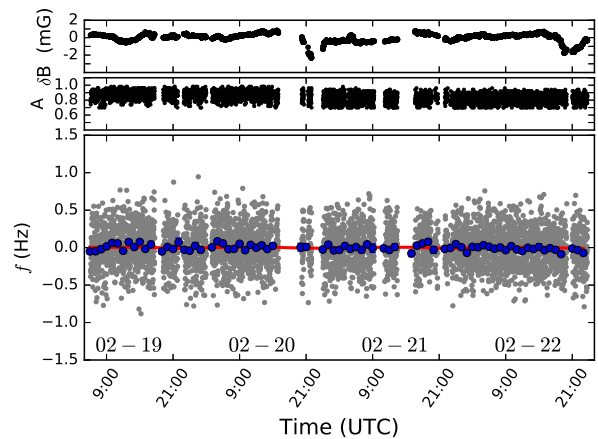}
\caption{ \label{fig:LLI_signal}  Magnetic field changes, $\delta B$, interference amplitude, and frequency measurements. The gray points represent the single measurement corrected for the quadratic Zeeman shift. We bin the measurements to 60 min intervals (blue points) and fit  to Eq. (\ref{Eq.tensot_fit}) (red curve) to bound the LLI tensor elements.   
}
\end{figure}

In total, one experimental block consists of measuring the time evolution for the states $\ket{\psi^{0,1}_R}$ and $\ket{\psi^{0,1}_L}$ for wait times of $\tau=\{5\:{\rm ms},\:105\:{\rm ms} \}$ and analysis phases of $\varphi = \{\varphi_0,\: (\varphi_0+90^\circ)\}$ (in total, eight measurements). In addition, we monitor the parity fringe amplitude for $\tau=5$~ms by using analysis laser phases of $\varphi=45^\circ$ and $\varphi=-45^\circ$, respectively.
Each measurement block lasts for about 40~s.
After each measurement block, we extract the phase and adjust the laser phase offset $\varphi_0$ so as to keep measuring near the zero crossing of the parity fringes.  To maintain high contrast, we insert calibration measurements of the carrier transition frequencies and magnetic field every 10 min. 
Following this procedure, we monitored the time evolution of the LLI state continuously from February 19, 2018 at 06:00 until February 23, 2018 at 03:00 coordinated universal time (UTC); see Fig.\ref{fig:LLI_signal}.

Additional contributions to the phase evolution arise from the quadratic Zeeman shift \cite{Sobelman1992} and
the interaction with the dc electric field gradient \cite{Roos2006}. 
The quadratic Zeeman shift was calculated to change the frequency by $4.5$~mHz per 1~mG change in the magnetic field  \cite{Sobelman1992}. 
The magnetic field was measured to change by less than than 4~mG during the experiment, see Fig.\ref{fig:LLI_signal}.
Using the measured magnetic fields, we applied a correction to the frequency to compensate for the small magnetic field changes. 
In an effort to reduce this energy shift, we carefully aligned the quantization angle with trap axis to ~$58^\circ$ \cite{Roos2006}. 
We measured the phase $\phi$ as a function of the axial frequency to accurately quantify the quadrupole shift.
We found that the quadrupole shift amounts to only $6.2$~Hz at an axial c.m. frequency of $\omega_{\rm c.m.}=830$~kHz, yielding a frequency shift of $-1.5$~mHz/kHz $\omega_{\rm c.m.}$. 
The axial c.m. frequency was measured to be stable to better than a $200$~Hz over 12~h and we can ignore the effect of these fluctuations on the LLI signal.

\begin{figure}
\includegraphics[angle=0,width=0.45\textwidth]{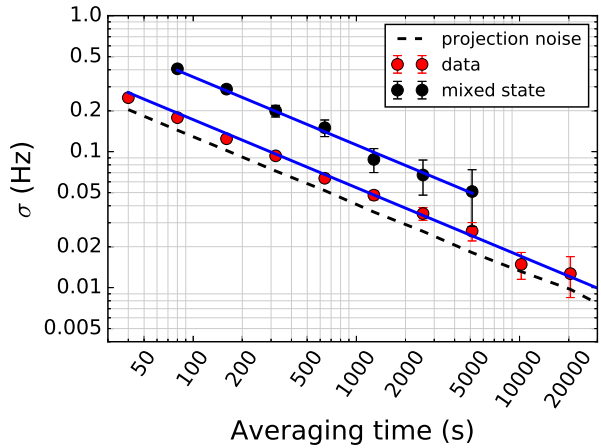}
\caption{
\label{fig:Allan_deviation} 
Allan deviation of the frequency measurements, $\sigma$ calculated from the unbinned data for an entangled state (red), and a mixed state prepared according to Ref.~\cite{Pruttivarasin2015} (black).
The blue solid line is a fit showing that $\sigma =1.72\: {\rm Hz}/ \sqrt{\tau} $  and $\sigma =3.54\: {\rm Hz}/ \sqrt{\tau} $ for the entangled and mixed states, respectively, where $\tau$ is the averaging time.
The dashed black line is the calculated projection noise for the entangled state. 
}
\end{figure}

We calculated the Allan deviation of the frequency measurement as a function of the averaging time $\tau$; see Fig.\ref{fig:Allan_deviation}. 
The Allan deviation is averaging as a function of $1/\sqrt(\tau)$,  indicating that we are still limited by statistical noise rather than by correlated noise or systematics as discussed above. 
The Allan deviation is also a measure of the uncertainty in the frequency estimate. We measure that, for the entangled state, the uncertainty decreases at a rate of $1.72\:{\rm Hz}/\sqrt{\tau}$, whereas for a separable state, it decreases as $3.54\: {\rm Hz}/ \sqrt{\tau}$. 
This is the expected improvement of a factor of two due to using entanglement over the previous sensitivity published in Ref~\cite{Pruttivarasin2015}.
In addition, our total measurement time was $\sim4$ times longer, leading to another improvement in the frequency uncertainty by a factor of two, resulting in a total frequency uncertainty of $3.4$~mHz.

We calculate the bounds for the LLI tensor coefficients, $C_{MN}$, from the measured frequency as given by Fig.~\ref{fig:LLI_signal}.
First, we use the Lorentz transformation to rotate LLI tensor coefficients from the laboratory reference frame, $C_{\mu'\nu'}$, to the SCCEF reference frame, $C_{MN}$.
For two ions, the energy difference due to a potential LLI violation is given by \cite{Hohensee2013c}
\begin{eqnarray} \label{Eq.tensot_fit}
    f=\nicefrac{\Delta E}{h} &=&  -8.9(2)\times10^{15}\:{\rm Hz}(C_{xx}+C_{yy}-2C_{zz}) \Rightarrow  \nonumber \\
     &=&  
    A {\rm sin}(\omega_\oplus T) +
    B {\rm cos}(\omega_\oplus T) + \nonumber \\
    &\quad& C {\rm sin}(2\omega_\oplus T) +
    D {\rm cos}(2\omega_\oplus T).
\end{eqnarray}
where $\omega_\oplus=2\pi/23.93$~h is the sidereal angular frequency of the Earth rotation, and $T$ is the time measured since the vernal equinox. 
The $A$,$B$,$C$,and $D$ coefficients depend on the $C_{MN}$, the colatitude angle of the experiment of $52.1^\circ$, and the angle of the magnetic field. 
We fit the hourly binned data in Fig.~\ref{fig:LLI_signal} to Eq.(\ref{Eq.tensot_fit}) and calculate the tensor coefficients, $C_{MN}$.
The results are summarized in Table \ref{tab:LLI_tensor}, where it is shown that we have set a new upper bound for any possible spatial violation of the local Lorentz invariance at about $5\times10^{-19}$. This represents a two- to fourfold improvement over the current bound.
 
 \begin{table}[] \label{Tab.LLI_params}
    \centering

\begin{tabular}{@{}|c|c|c|c@{}|}
\toprule
\hline
Parameter &  & \multicolumn{1}{c|}{New limit} & \multicolumn{1}{c|}{Existing limit \cite{Pruttivarasin2015}} \\ \hline \midrule
$C_{X-Y}$  &  & $(6.2\pm 9.2)\times10^{-19}$                      & $(0.2\pm 2.3)\times10^{-18}$    \\

$C_{XY}$   &  & $(2.4\pm 4.8)\times10^{-19}$                      & $(-0.8\pm 1.2)\times10^{-18}$     \\

$C_{XZ}$   &  & $(0.8\pm 2.1)\times10^{-19}$                      & $(3.4\pm 7.9)\times10^{-19}$     \\

$C_{YZ}$   &  & $(-3.1\pm 2.2)\times10^{-19}$                     & $(1.7\pm 7.1)\times10^{-19}$      \\ \hline \bottomrule
\end{tabular}
    \caption{
    Limits on Lorentz violation parameters (in the SCCEF) given by fitting our data to the model in Eq.~\ref{Eq.tensot_fit}.  
    The uncertainties are one standard deviation from the fit scaled by the calculated $\sqrt{\chi^2_{red}}=1.17$. 
    Note that we use the notation $C_{X-Y}=C_{XX}-C_{YY}$.
    }
    \label{tab:LLI_tensor}
\end{table}



One may wonder about the additional gain when scaling to Greenberger–Horne–Zeilinger states with more particles. For instance, states of the form $\ket{(D_\frac{5}{2}D_{-\frac{5}{2}}D_{-\frac{5}{2}}D_\frac{5}{2}+D_\frac{1}{2}D_{-\frac{1}{2}}D_{-\frac{1}{2}}D_\frac{1}{2})}$ can be generated with the exact same pulse sequence as used in this work and, in addition to the global magnetic field fluctuations,  decouple the magnetic field gradient. This may become an important consideration when applying our scheme to states with ultralong lifetimes, such as to the F$_{7/2}$ state for Yb$^+$ ions \cite{Dzuba2016a}. In this case,  using the four-ion entangled state above improves the signal-to-noise ratio by a factor of $2^4/2=8$  as compared to preparing it probabilistically via the separable state. 
We test this approach and create a four-ion entangled state. 
However, for \Ca, spontaneous emission limits the coherence before the magnetic field gradient does. 
Because spontaneous emission acts on each ion independently, the coherence time of the four-ion state is halved as compared to the two-ion state. 
Thus, one expects the same  signal-to-noise ratio as for two uncorrelated two-ion states, i.e., only an improvement of $1/\sqrt{2}$ as compared to our experiments above.
We were able to achieve an entangling gate fidelity of $\sim$80\%, which resulted in a LLI-state preparation of $\sim$50\%.
In addition,
complications such as an increased time overhead for cooling and increased sensitivity to infidelities of the single-qubit rotations lead to an actual decrease of the sensitivity.
All of these complications are not of a fundamental nature, and our measurements show that using more complicated entangled states is indeed a viable route. Nevertheless, these measurements also illustrate that taking advantage of entanglement requires a high degree of experimental control.

In conclusion, we use tailored quantum correlations to eliminate the first-order sensitivity to fluctuations in the global magnetic field the dominant decoherence mechanism in our experiment. 
Furthermore, we increase our measurement signal as compared to similar experiments (Refs.~\cite{Pruttivarasin2015,Chwalla2007}) performed using mixed states.
We show that, through the use of high-fidelity entangling operations, the signal can be improved by nearly a factor of two, which is near the projection noise limit.
We have applied this method to improve the bounds on spatial violation of the local Lorentz symmetry of the electron to about $5\times10^{-19}$. 
Our measurements demonstrate empirically that entanglement can be used to improve precision measurements. In particular, this is the case if quantum states can be engineered which  do not couple to the dominant noise sources but are still sensitive to the signal. Although, in principle, the gains are expected to be exponential in the number of entangled parties as compared to using separable states, the scaling of the required resources and alternative measurement schemes needs to be assessed carefully.

\begin{acknowledgments}
We thank Wei-Ting Chen and Ryan Shaffer for help with data taking.
This work has been supported by the NSF through Grant No. PHY 1507160 and the AFOSR through Grant No.~FA9550-15-1-0249. 
\end{acknowledgments}

\bibliographystyle{apsrev4-1}
\bibliography{mendeley}

\end{document}